\documentclass[aps,twocolumn,groupedaddress,showpacs]{revtex4}
\usepackage{graphicx,color}
\begin{document}

\title{Effect of the cosmological constant on the bending  of light\\
and the cosmological lens equation}

\author{Hideyoshi ARAKIDA}
\email[E-mail:]{arakida@iwate-u.ac.jp}
\affiliation{Graduate School of Education, Iwate University, 
Morioka, Iwate 020-8550,Japan}

\author{Masumi KASAI}
\email[E-mail:]{kasai@phys.hirosaki-u.ac.jp}
\affiliation{Graduate School of Science and Technology,
Hirosaki University, 
Hirosaki, Aomori 036-8561 Japan}

\date{\today}

\begin{abstract}
  We revisit the effect of cosmological constant $\Lambda$ on the
  light deflection and its role in the cosmological lens equation.
  First, we re-examine the motion of photon in the Schwarzschild
  spacetime, and explicitly describe the trajectory of photon and
  deflection angle $\alpha$ up to the second-order in $G$. Then the
  discussion is extended to the contribution of the cosmological
  constant $\Lambda$ in the Schwarzschild-de Sitter or Kottler
  spacetime. Contrary to the previous arguments, we emphasize the
  following points: (a) the cosmological constant $\Lambda$ does
  appear in the orbital equation of light, (b) nevertheless the
  bending angle of light $\alpha$ does not change its form even if
  $\Lambda \neq 0$ since the contribution of $\Lambda$ is thoroughly
  absorbed into the definition of the impact parameter, and (c) the
  effect of $\Lambda$ is completely involved in the angular diameter
  distance $D_A$.
\end{abstract}

\maketitle

\section{Introduction}

Nowadays, it is widely regarded that the cosmological constant
$\Lambda$ or more generally dark energy is the most responsible
candidate which explains accelerating expansion of
Universe. Nonetheless the details of cosmological constant $\Lambda$
or dark energy 
are still far from clear, then it is preferable and
worthy to clarify the validity of this hypothesis by means of not only
cosmological observations but also another astronomical/astrophysical
ones.

Among such attempts, it would be the most natural idea to investigate
the role of cosmological constant $\Lambda$ in the classical tests of
general relativity, e.g. the perihelion advance of planetary orbit and
the bending of light path. So far, it was shown that the cosmological
constant $\Lambda$ causes the perihelion shift of planets at least in
principle, even though its contribution is too small to detect in the
current measurement technique (see
\cite{kerr2003,rindler2006,iorio2008} and 
the references therein).

While it has been believed for a long time that $\Lambda$ does not
contribute to the light deflection because there is no $\Lambda$ in
the second-order ordinary differential equation (ODE) of
photon. However recently, Rindler and Ishak \cite{rindler2007} pointed
out that $\Lambda$ does affect the bending angle by using the
Schwarzschild-de Sitter or Kottler metric and the invariant formula of
cosine. Subsequently many authors argued its appearance in diverse
ways and generality assisted the fact that there appears $\Lambda$ in
the deflection angle $\alpha$, see \cite{ishak2010} for review and
the references therein and also
\cite{lake2002,park2008,kp2008,sph2010,bhadra2010,miraghaei2010,biressa2011}. 
However, it seems that the conclusion has not converged yet; 
for instance whether the leading order effect of $\Lambda$ is coupled 
with the mass of central body $M$ or not and so on.
In order to clear up the confusion, we
will revisit the effect of the cosmological constant on the light
deflection and its role in the cosmological lens equation.


\section{Photon trajectory in Schwarzschild spacetime\label{light_def1}}

Before discussing the influence of $\Lambda$ on bending angle $\alpha$, 
we shall begin with re-considering the solution of photon trajectory 
in the Schwarzschild spacetime. From the Schwarzschild metric in 
the Schwarzschild coordinates,
\begin{eqnarray}\label{eq:1}
 ds^2 = - \left(1 - \frac{r_g}{r}\right)c^2dt^2 +
  \left(1 - \frac{r_g}{r}\right)^{-1}dr^2   \nonumber \\
 \quad + r^2(d\theta^2 + \sin^2\theta d\phi^2),\ \ 
  r_g = \frac{2GM}{c^2},
\end{eqnarray}
and the condition for null geodesic $ds^2 = 0$, we have the
geodesic equation for equatorial plane ($\theta = \pi/2$),
\begin{eqnarray}
\left(\frac{du}{d\phi}\right)^2 = \frac{1}{b^2} - 
 u^2 + r_g u^3,\quad
 u \equiv \frac{1}{r},\quad \frac{1}{b^2} \equiv 
 \frac{E^2}{c^2L^2},
 \label{eq:geodeq1}
\end{eqnarray}
in which the two constants of motion, $E$ and $L$ are total 
energy and angular momentum, respectively. Alternatively, 
Eq.~(\ref{eq:geodeq1}) can be expressed in the form of 
second-order ODE as,
\begin{eqnarray}
 \frac{d^2 u}{d\phi^2} = -u + \frac{3}{2}r_g u^2,
  \label{eq:geodeq2}
\end{eqnarray}
nevertheless, hereinafter we use Eq.~(\ref{eq:geodeq1}) instead of 
Eq.~(\ref{eq:geodeq2}).

In order to obtain the photon trajectory up to the second-order
in $G$, let us put the solution of Eq.~(\ref{eq:geodeq1}) $u$ as,
\begin{eqnarray}
 u = \frac{1}{b}\left(\sin \phi + r_g u_1 + r^2_g u_2\right),
\end{eqnarray}
where $u_1$ and $u_2$ are, respectively, 
first ${\cal O}(G)$ and second  
${\cal O}(G^2)$ order correction to zeroth-order solution 
$u_0 = \sin \phi/b$ (straight line). Hence $u_1$ and $u_2$ satisfy
the following differential equations,
\begin{eqnarray}
 \frac{du_1}{d\phi} &=& - \frac{\sin\phi}{\cos\phi} u_1 
 + \frac{1}{2b}\frac{\sin^3\phi}{\cos\phi},
 \label{eq:1st-eq}\\
 \frac{du_2}{d\phi} &=& - \frac{\sin\phi}{\cos\phi} u_2 
  \label{eq:2nd-eq} \\
&& \quad - \frac{1}{2\cos\phi}
 \left[\left(\frac{du_1}{d\phi}\right)^2 + u_1^2
 - \frac{3}{b}u_1 \sin^2\phi\right].
 \nonumber
\end{eqnarray}
Noting that the integration constants of Eqs.~(\ref{eq:1st-eq}) 
and (\ref{eq:2nd-eq}) are chosen such that maximizing $u$ 
(or minimizing $r$) for $\phi = \pi/2$, then we obtain the trajectory 
of photon up to the second-order in $G$ as,
\begin{eqnarray}
 \frac{1}{r} &=& 
  \frac{1}{b}\sin\phi + 
  \frac{r_g}{4 b^2}\left(3 + \cos 2\phi\right)
    \label{eq:trajectory1} \\ 
 & & 
  + \frac{r_g^2}{64 b^3}
  \bigl(37\sin\phi + 30\left(\pi-2\phi\right)\cos\phi - 3\sin
   3\phi\bigr), \nonumber
\end{eqnarray}
where $b$ is the impact parameter which represents the
minimum value of $r$-coordinate for the undeflected light ray, i.e.,
$r_g=0$.
The bending angle $\alpha$ is shortly derived from
Eq.~(\ref{eq:trajectory1}) and it coincides with the famous formula by
\cite{epstein1980},
\begin{eqnarray}
 \alpha &=& 
 2\frac{r_g}{b}+ 
\frac{15\pi}{16}\left(\frac{r_g}{b}\right)^2
\nonumber \\ 
&=& \frac{4GM}{c^2 b} + \frac{15\pi}{4}\frac{(GM)^2}{c^4 b^2}
  + {\cal O}(G^3).
  \label{eq:def_angle1}
\end{eqnarray}
A simple derivation of $\alpha$ is given in Appendix \ref{B}.
It should be mentioned about the validity of the solution for
light trajectory. The appropriateness of our solution, 
Eq.~(\ref{eq:trajectory1}) can be verified readily by the direct 
substitution into Eq.~(\ref{eq:geodeq1}) and it is found that the 
residual terms are order ${\cal O}{(G^3)}$, then it is perfectly valid 
up to the order of $G^2$. However, the photon trajectory given in 
previous works such as Eq.~(18) of \cite{bodenner2003} and
Eq.~(16) of \cite{ishak2010} are incorrect; in fact, there appears 
${\cal O}(G^2)$ order residual term in the solution of photon trajectory 
in \cite{bodenner2003,ishak2010}.

\section{Contribution of the cosmological constant\label{light_def2}}

Now, let us investigate the contribution of $\Lambda$ on light ray. 
For this purpose, we adopt the Schwarzschild-de 
Sitter or Kottler metric \cite{kottler1918},
\begin{eqnarray}\label{eq:9}
 ds^2 &=& - \left(1 - \frac{r_g}{r} - \frac{\Lambda}{3}r^2\right)c^2dt^2 
\\ \nonumber
&& +
  \left(1 - \frac{r_g}{r} - \frac{\Lambda}{3}r^2\right)^{-1}dr^2  
+ 
  r^2(d\theta^2 + \sin^2\theta d\phi^2).
\end{eqnarray}
In the same way as the Schwarzschild case, the differential equation 
of light is given by
\begin{eqnarray}
\left(\frac{du}{d\phi}\right)^2 = \frac{1}{b^2} - 
 u^2 + r_g u^3 + \frac{\Lambda}{3}.
 \label{eq:geodeq3}
\end{eqnarray}
It should be emphasized here that the geodesic equation of light 
Eq.~(\ref{eq:geodeq3}) does obviously include $\Lambda$.  
Therefore, previous arguments, such as 
``$\Lambda$ does not appear in the geodesic equation of light'',
would be overstated. Actually, the second-order ODE derived from 
Eq.~(\ref{eq:geodeq3}) reduces to Eq.~(\ref{eq:geodeq2}), 
nevertheless its solution of light trajectory 
should be obtained in such a way that the integration constants 
satisfy Eq.~(\ref{eq:geodeq3}).

Furthermore, the impact parameter is 
the minimum value of the coordinate $r$
if the light ray were undeflected, 
i.e., $r_g=0$. It is obvious from Eq.~(\ref{eq:geodeq3}) that 
the impact parameter $B$ in this case is defined by
\begin{eqnarray}
 \frac{1}{B^2} \equiv \frac{1}{b^2} + \frac{\Lambda}{3}. 
  \label{eq:impact}
\end{eqnarray}
Then, the form of Eq.~(\ref{eq:geodeq3}) completely coincides with 
Eq.~(\ref{eq:geodeq1}), except that the impact parameter $b$ is 
replaced by $B$.
Therefore, the solution of Eq.~(\ref{eq:geodeq3}) becomes,
\begin{eqnarray}
\frac{1}{r} &=& 
  \frac{1}{B}\sin\phi + 
  \frac{r_g}{4 B^2}\left(3 + \cos 2\phi\right)
    \label{eq:trajectory2}\\
 & & 
  + \frac{r_g^2}{64 B^3}
  \bigl(37\sin\phi + 30(\pi-2\phi)\cos\phi - 3\sin
   3\phi\bigr),
   \nonumber
\end{eqnarray}
and deflection angle is,
\begin{eqnarray}
 \alpha &=& 
  2\frac{r_g}{B}+ 
  \frac{15\pi}{16}\left(\frac{r_g}{B}\right)^2 \nonumber\\
  &=& \frac{4GM}{c^2 B} + \frac{15\pi}{4}\frac{(GM)^2}{c^4 B^2}
  + {\cal O}(G^3).
  \label{eq:def_angle2}
\end{eqnarray}
It is worthy to note that the contribution of $\Lambda$ is 
incorporated in Eqs.~(\ref{eq:trajectory2}) and (\ref{eq:def_angle2})
through Eq.~(\ref{eq:impact}). As a consequence, it is 
found that the cosmological constant $\Lambda$ does appears in 
both the geodesic equation and its solution, that is the trajectory 
of photon. 
However, the effect of $\Lambda$ is completely 
absorbed into the definition of the the impact parameter
 (see Eq.~(\ref{eq:impact})).
Hence it is difficult
to distinguish the influence of $\Lambda$ from the observed 
deflection angle.

When we expand Eq.~(\ref{eq:def_angle2}) by using
$1/B = (1/b)\sqrt{1 + \Lambda b^2/3} \simeq (1/b)(1 
+ \Lambda b^2/6)$
and remain ${\cal O}(M, M\Lambda)$ terms, it follows that
\begin{eqnarray}
 \alpha \simeq \frac{4GM}{c^2 b} + \frac{2GMb\Lambda}{3c^2},
  \label{eq:def_angle3}
\end{eqnarray}
in which second term coincides with the previous results, e.g.
Eq.~(5) and below in \cite{sereno2009} and the third term of Eq.~(15)
in \cite{bhadra2010}. Hence it is found that these results are
included in Eq.~(\ref{eq:def_angle2}) as a limiting case.

It is clear that the trajectory of photon Eq.~(\ref{eq:trajectory2})
strictly satisfies Eq.~(\ref{eq:geodeq3}) up to the second-order 
in $G$ based on the result in previous section.

\section{Cosmological lens equation\label{lens}}

Finally, we consider the contribution of $\Lambda$ in the cosmological
lens equation. 
Under the assumption that the thin lens approximation 
is valid, the lens equation relates the observed image position angle
$\theta$ to the unlensed position angle $\beta$ of the source as
\begin{equation}
  \beta = \theta - \frac{D_A(z_L,z_S)}{D_A(0,z_S)} \alpha, 
\label{eq:lenseq1}
\end{equation}
where $D_A(z_1,z_2)$ denotes the angular diameter distance from the
redshift $z_1$ to $z_2$, and the arguments $z_L$ and $z_S$ are the
redshift of the lens and the source, respectively. 
For the distance formula $D_A$ in the unperturbed
Friedmann-Lema\^itre-Robertson-Walker (FLRW) universe with $\Lambda$, 
see, e.g., \cite{FFKT}.

Up to the first order in $G$, the bending angle $\alpha$ in the
present case is
\begin{equation}
  \alpha = \frac{4GM}{c^2 B}.
   \label{eq:lensangle}
\end{equation}
The impact parameter $B$ is related to the image position angle
$\theta$ by
\begin{eqnarray}
 B = D_A(0,z_L) \theta. 
\label{eq:cosmolens}
\end{eqnarray}
Then, 
from Eqs.~(\ref{eq:lenseq1}), (\ref{eq:lensangle}), and 
(\ref{eq:cosmolens}), the lens equation is finally
\begin{equation}\label{eq:lenseq2}
  \beta = \theta - \frac{4GM D_A(z_L,z_S)}{c^2 D_A(0,z_L)\,
    D_A(0,z_S)}\frac{1}{\theta} . 
\end{equation}
Therefore, the contribution of $\Lambda$ is completely involved in the
form of the angular diameter distance $D_A$. No modifications due to
$\Lambda$ appear even if  the term of ${\cal O}(G^2)$ in $\alpha$ is included. 
It should also be noted that Eq.~(\ref{eq:lenseq2}) is exactly the same
form appeared in \cite{FFKT,FFK}, where the authors have shown that the
gravitational lensing effects are strongly dependent on the value of
the cosmological constant and hence they provide us with useful means
to test the cosmological constant.

Here, we mention that in paper \cite{sereno2009}, Sereno introduced
the relation, $b_0 = D_d \vartheta$ (see Eq.~(8) below of 
\cite{sereno2009}) where $b_0$ is the impact factor of 
``Schwarzschild lens'' (see Eq.~(8) of \cite{sereno2009}), 
$D_d$ is the angular-diameter distance, and $\vartheta$ is the
angular separation. Since in the case of cosmological lens, 
the cosmological distance, such as the angular diameter distance, 
is defined with $\Lambda$, then $b_0$ should be replaced by $B$, 
instead.

\section{Summary\label{summary}}

We briefly summarize our conclusions. 
Contrary to the previous 
arguments,  (a) the cosmological constant
$\Lambda$ does appear in the orbital
equation of light, (b) nevertheless the bending angle of light $\alpha$
does not change its form even if $\Lambda \neq 0$ since the contribution
of $\Lambda$ is thoroughly absorbed into the definition of the
impact parameter, and (c) the effect of $\Lambda$ is completely 
involved in the angular diameter distance $D_A$.  Then, no
modifications due to $\Lambda$ appear even if the second-order
terms in $G$ are included.

It should be mentioned that the similar conclusions 
are reached by other authors \cite{lake2002,park2008,kp2008,sph2010}.
We hope that this paper provides a simpler and clearer 
explanation of the role of $\Lambda$ in the bending of light and 
the cosmological lens equation than the above-mentioned studies.

\appendix

\section{The angle between the radial direction and the light
  trajectory\label{A}} 

Rindler and Ishak \cite{rindler2007} have derived 
the angle between the radial direction and the light
trajectory from the invariant cosine formula. 
Just for a reference, here we give a more intuitive derivation. 
Consider a spherically symmetric space.  The metric on
the equatorial plane $\theta=\pi/2$ is given by
\begin{equation}\label{eqa1}
  d\ell^2 = \frac{dr^2}{f(r)} + r^2 d\phi^2. 
\end{equation}
In this space, the infinitesimal proper lengths along the radial
direction and that perpendicular to the radial direction are given by 
 $f(r)^{-1/2} |dr|$ and $r|d\phi|$, respectively. 
Then the tangent of the angle $\psi$  between the radial direction and the light
  trajectory is
  \begin{equation}\label{eqa2}
    \tan\psi = \frac{r |d\phi|}{f(r)^{-1/2} |dr| } =
    \sqrt{f(r)}\,r \left|\frac{dr}{d\phi}\right|^{-1}, 
  \end{equation}
which is the same form as Eq.~(16) in \cite{rindler2007}. 
Once we obtain the solution $r(\phi)$ for the light trajectory,
Eq.~(\ref{eqa2}) gives the angle at any point $\left(r(\phi),
  \phi)\right)$ on the trajectory.  

\begin{figure}[h]
  \centering
  \includegraphics[width=0.42\textwidth]{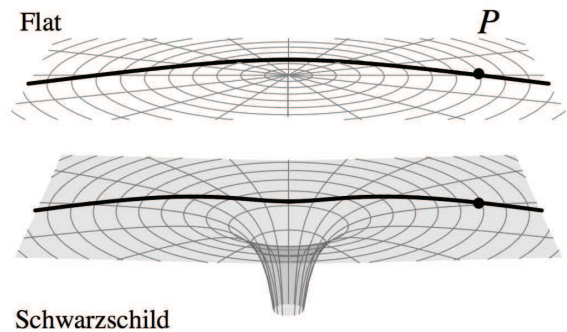}
\caption{Light trajectories on Schwarzschild geometry and on the flat
  space which is tangent to the asymptotically flat region of the
  Schwarzschild geometry.  The coordinates of the point $P$ is $\left(r(\phi),
  \phi)\right)=\left( r(0), 0 \right)$. \label{fig1}}
\end{figure}
\section{The bending angle in Schwarzschild geometry\label{B}}

\begin{figure}[ht]
  \centering
  \includegraphics[width=0.42\textwidth]{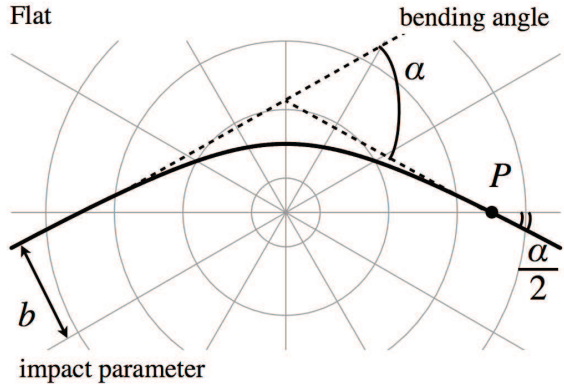}
\caption{Light trajectory on the flat space with the metric
  $d\ell^2=dr^2 + r^2d\phi^2$. The impact parameter $b$ and the
  bending angle $\alpha$ are defined on this space.\label{fig2} }
\end{figure}

Using Eqs.~(\ref{eq:1}), (\ref{eq:trajectory1}) and (\ref{eqa2}), the
angle $\psi$ at the point $P(r(0), 0)$ is
\begin{equation}
  \psi \simeq \sqrt{f(r(0))}\,r(0)
  \left|\frac{dr}{d\phi}\right|^{-1}_{\phi=0}
  \simeq \frac{r_g}{b}+
  \frac{15\pi}{32}\left(\frac{r_g}{b}\right)^2. 
\end{equation}
Consider a flat space which is tangent to the asymptotically
flat region of the Schwarzschild geometry (see Fig.~\ref{fig1}).  It
is apparent from the 
property of an isosceles triangle in the Euclidean geometry that the
angle $\psi$ at the point $P$ is a half of the bending angle $\alpha$
(see Fig.~\ref{fig2}).
Then,
\begin{equation}
  \alpha = 2\psi \simeq 2\frac{r_g}{b}+
  \frac{15\pi}{16}\left(\frac{r_g}{b}\right)^2. 
\end{equation}

\begin{figure}[h]
  \centering
  \includegraphics[width=0.42\textwidth]{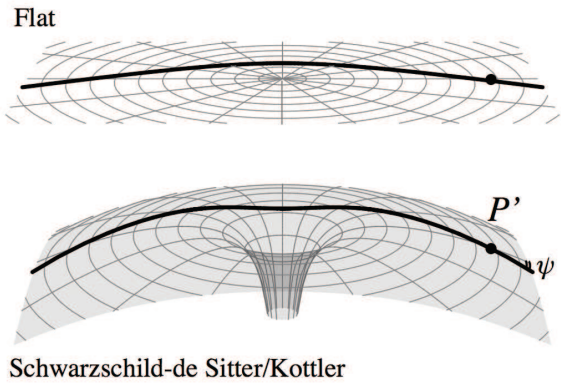}
\caption{The angle $\psi$ between the radial direction and the light
  trajectory at $P'$ on Schwarzschild-de Sitter geometry.\label{fig3}}
\end{figure}
\section{The angle on Schwarzschild-de Sitter geometry}

Using Eqs.~(\ref{eq:9}), (\ref{eq:trajectory2}), and (\ref{eqa2}), the
angle $\psi$ between the radial direction and the light trajectory at
the point $P'\left(r(0),0\right)$ on the Schwarzschild-de Sitter
  geometry is (see Fig. \ref{fig3})
  \begin{equation}\label{eqc1}
    \psi \simeq
    \sqrt{1-\frac{\Lambda}{3}\left(\frac{b^2}{r_g}\right)^2}
    \left( \frac{r_g}{b} + 
\frac{15\pi}{32}\left(\frac{r_g}{b}\right)^2\right). 
  \end{equation}
When linearized with respect to $\Lambda$, 
\begin{equation}\label{eqc2}
     \psi \simeq
\frac{r_g}{b}+
  \frac{15\pi}{32}\left(\frac{r_g}{b}\right)^2
-\frac{\Lambda}{6}\frac{b^3}{r_g} - \frac{15\pi}{192}\Lambda b^2, 
\end{equation}
which is essentially a half of Eq.~(18) in \cite{ishak2010}, although
they drop the last term in Eq.~(\ref{eqc2}). 
It should be emphasized since the Euclidean
property does not 
hold on the Schwarzschild-de Sitter geometry, the angle $\psi$ in 
Eq.~(\ref{eqc2}) is {\it not} a half of the bending angle
$\alpha$.

\end{document}